\documentclass{llncs}

\newcommand{\ehp}{\textsc{Edge Hamiltonian Path}}
\newcommand{\ehc}{\textsc{Edge Hamiltonian Cycle}}
\newcommand{\des}{\textsc{Dominating Eulerian Subgraph}}
\newcommand{\epath}{edge-Hamiltonian path}
\newcommand{\ecycle}{edge-Hamiltonian cycle}
\newcommand{\egraph}{edge-Hamiltonian graph}
\newcommand{\dgraph}{dominating Eulerian subgraph}

\begin{document}

\title{Parameterized Edge Hamiltonicity}
\author{Michael Lampis\inst{1,}\thanks{Research supported by a Scientific
Grant-in-Aid  from the  Ministry of Education, Culture, 
Sports, Science and Technology of Japan.}, Kazuhisa Makino\inst{1}, Valia Mitsou\inst{2,}\thanks{This work was done while the author was visiting RIMS, Kyoto University.}, Yushi Uno\inst{3}}
\institute{Research Institute for Mathematical Sciences, Kyoto University \\\email{mlampis,makino@kurims.kyoto-u.ac.jp} \and CUNY Graduate Center \\\email{vmitsou@gc.cuny.edu} \and Department of Mathematics and Information Sciences, Graduate School of Science, Osaka Prefecture University \\\email{uno@mi.s.osakafu-u.ac.jp}}

\maketitle

\begin{abstract}

We study the parameterized complexity of the classical \ehp\ problem and give
several fixed-parameter tractability results.  First, we settle an open
question of Demaine et al.~by showing that \ehp\ is FPT parameterized by vertex
cover, and that it also admits a cubic kernel.  We then show fixed-parameter
tractability even for a generalization of the problem to arbitrary hypergraphs,
parameterized by the size of a (supplied) hitting set. We also consider the
problem parameterized by treewidth or clique-width. Surprisingly, we show that
the problem is FPT for both of these standard parameters, in contrast to its
vertex version, which is W-hard for clique-width.  Our technique, which may be
of independent interest, relies on a structural characterization of
clique-width in terms of treewidth and complete bipartite subgraphs due to Gurski
and Wanke.

\end{abstract}

\section{Introduction}

The focus of this paper is the \ehp\ problem, which can be defined as follows:
given an undirected graph $G(V,E)$, does there exist a permutation of $E$ such
that every two consecutive edges in the permutation share an endpoint? This is
a very well-known graph-theoretic problem, which corresponds to the restriction
of (vertex) \textsc{Hamiltonian Path} to line graphs. Despite some superficial
similarity to the problem of finding an Eulerian path, this problem has long
been known to be NP-complete, even for graphs which are bipartite or have
maximum degree 3 \cite{B81,RWX11,LW93}.

The \ehp\ problem is a very natural graph-theoretic problem with a long
history (see e.g.~\cite{BS81,C92,C84,C68,L98,CLL03}). 
In this paper we investigate the complexity of this problem from the
parameterized complexity perspective. More specifically, we consider the case
where some structural parameter of the input graph $G$, such as its treewidth,
has a moderate value. Despite the problem's prominence, to the best of our
knowledge, \ehp\ has never before been studied in this setting. Such an
investigation is  of inherent interest from the point of view of graph theory
and parameterized complexity. Beyond this, we are partially motivated by a
specific question recently asked explicitly by Demaine et al.~\cite{DDH14}.  In
their investigation of the card game UNO, the authors of \cite{DDH14} present
an XP (i.e.~running in $n^{f(k)}$) dynamic programming algorithm for \ehp\ on
bipartite graphs, where $k$ is the size of the smaller part (that is, $k$ is
the size of a vertex cover).  They then, quite naturally, ask if this can be
improved to an FPT algorithm. In this paper we present a number of results that
positively settle not only this, but several other more general such questions.

\noindent\textbf{Overview of results} We give fixed-parameter tractability
results for \ehp\ and its variant \ehc, which we show to be essentially
equivalent.  Our first task is to consider the problem parameterized by the
size of the vertex cover of the input graph. We establish that, not only is the
problem FPT, but it also admits a cubic kernel through an algorithm that
locates and deletes irrelevant edges.  We then go on to give a much more
general direct FPT algorithm for the problem and show that this algorithm can
still be applied even if we consider the problem on arbitrary hypergraphs and
the parameter is the size of a hitting set which is supplied with the input.
As a corollary, we note that this result implies that (vertex)
\textsc{Hamiltonian Path} is FPT when parameterized by the chromatic number of
the complement of the input graph.

Our next direction is to consider the  problem on graphs parameterized by
treewidth and clique-width. The complexity of \ehp\ for these parameters was
previously unknown, since this is also a more general question than the one
posed in \cite{DDH14}. Our first observation is that fixed-parameter
tractability for \ehc\ parameterized by treewidth can be obtained from standard
meta-theorems, if one relies on an alternative characterization of the problem.
This alternative characterization, which was first given by Harary and
Nash-Williams almost 50 years ago \cite{HN65}, allows one to recast this
ordering problem as the problem of finding an appropriate Eulerian subgraph,
which, with a little work, can be expressed in a variant of Monadic Second
Order logic. For the sake of completeness, we also sketch a direct
treewidth-based dynamic programming algorithm using this formulation.

Having settled the problem for treewidth, the natural next step is to consider
\ehc\ parameterized by clique-width, a prominent structural graph parameter
that generalizes treewidth. It is important to note here that the (more common)
vertex version of the problem exhibits a sharp complexity jump between these
two parameters: \textsc{Hamiltonian Cycle} is FPT for treewidth but for
clique-width the problem is W-hard and therefore does not admit an FPT
algorithm under standard complexity assumptions \cite{FGLS09}.  In what is
perhaps the most surprising result of this paper, we show that \ehc\ remains
FPT even for clique-width, despite this parameter's additional generality.  On
a high level, our strategy is to rely on a characterization of bounded
clique-width graphs given by Gurski and Wanke \cite{GW00} which states roughly
that if a graph has small clique-width and no large complete bipartite
subgraphs, then it has small treewidth. We devise an algorithm that locates and
``reduces'' large complete bipartite subgraphs in the input graph, without
affecting the answer or increasing the clique-width.  By repeatedly applying
this step we end up with a graph of small treewidth for which the problem is
FPT. We believe this idea, which is a rare algorithmic application of the
characterization of \cite{GW00}, may be of independent interest.

\section{Preliminaries}

We assume that the reader is familiar with the basics of parameterized
complexity. In particular, we use the definitions of the classes FPT, XP as
well as the notion of a kernelization algorithm and of polynomial kernels (see
\cite{DF99,FG06,N06}).

We will use the definition of treewidth, and in particular the notion of
``nice'' tree decompositions (see the survey \cite{BK08}). We also use the
notion of clique-width (see \cite{EGW01,CMR00,HOS08}). Let us briefly review
the definition. The class of graphs of clique-width $k$ contains all
single-vertex graphs where the only vertex has a label from $\{1,\ldots,k\}$.
Furthermore, the class is closed under the following operations: disjoint union
of two graphs; renaming of all vertices with some label $i$ to some label $j$;
and joining by new edges of all vertices with some label $i$ to all vertices
with some label $j$. When given a graph of clique-width $k$ we assume, as is
customary, that we are also supplied a clique-width expression, that is, a
rooted binary tree showing how the graph can be obtained from single-vertex
graphs using the above operations. All graph classes with bounded treewidth
also have bounded clique-width, but the reverse is not true \cite{CR05}.

We will also rely on the following theorem of Gurski and Wanke which
intuitively states that large complete bipartite graphs are what separates
treewidth from clique-width:

\begin{theorem}[\cite{GW00}] \label{thm:gw}

Let $G$ be a graph of clique-width $k$. If $G$ does not contain the complete
bipartite graph $K_{t,t}$ as a subgraph, then the treewidth of $G$ is at most
$3kt$.

\end{theorem}

We will consider the \ehp\ and \ehc\ problems. As mentioned, in these problems
we are looking for a permutation of the edges of the input graph so that any
two consecutive edges share an endpoint (in the latter problem, also the first
and last edge must share an endpoint). We call such a permutation an \epath\
(respectively an \ecycle). We will mostly view these as graph problems, but
this problem definition applies equally well to hypergraphs, if we require that
two consecutive \emph{hyperedges} share a common vertex.  Hypergraphs are the
subject of Section \ref{sec:hyper}. Recall that for a graph or hypergraph
$G(V,E)$, its line graph is the graph $G'(E,H)$ where $(e_1,e_2)\in H$ if and
only if $e_1,e_2$ share a vertex in $G$. The \ehp\ problem on $G$ is equivalent
to the \textsc{Hamiltonian Path} problem on $G'$.

For the graph case, it will be useful to recast these ordering problems as
subgraph problems.  First, recall that a graph is Eulerian if it is connected
and all its vertices have even degree. A \des\ of a graph $G(V,E)$ is a
subgraph $G'(V',E')$ of $G$ such that all edges of $E$ have an endpoint in
$V'$, that is, $V'$ is a vertex cover of $G$, and $G'$ is Eulerian.  We will
use the following classical observation of Harary and Nash-Williams:

\begin{theorem}[\cite{HN65}] \label{thm:hn}

A graph has an \ecycle\ if and only if it contains a \dgraph.

\end{theorem}

Finally, let us mention that we will deal with \ehp\ and \ehc\ interchangeably,
depending on which problem makes the description of our algorithms easier. The
reader can easily verify that all our arguments apply to both problems with
very minor modifications. It is also not hard to show the following:

\begin{lemma} \label{lem:equiv}

For the following parameters, \ehp\ is FPT if and only if  \ehc\ is FPT: vertex
cover, treewidth, clique-width and hypergraph hitting set.

\end{lemma}

\section{Vertex Cover} \label{vckernel}

In this section we consider the \ehp\ problem parameterized by the size of the
vertex cover $k$. We show that the problem has a cubic in $k$ kernel. As in the
following sections, we assume that together with the input graph $G(V,E)$ we
are given a vertex cover $S$ of $G$ with $|S|=k$. Note though, that this
assumption is not important, since a 2-approximate vertex cover can be found in
polynomial time.

Below follow some definitions which will make the presentation of the results
smoother.  We assume that the vertices of $G$ are labeled in some
lexicographically ordered fashion, and in particular that
$S=\{u_1,\ldots,u_k\}$.

\begin{definition}\label{def:type} 

An edge $e\in E$ is defined to be of \emph{type $i$} if it is incident to
$u_i\in S$ but not incident to any other $u_j\in S$ for $j<i$.  

\end{definition}

\begin{definition}\label{def:group} 

Let $P$ be an \epath\ of $G$.  For $i\in \{1,\ldots,k\}$,  a \emph{group of
type $i$} is a maximal set of edges of type $i$ which are consecutive in $P$.
We say that an edge is \emph{special} if it is the first or the last edge of a
group.
 
\end{definition}
 
The special edges essentially form the backbone of the \epath\ $P$.  A piece of
intuition that will become useful later is that, if one fixes these edges in a
proper edge-path, the remaining edges will be easy to deal with, because they
are allowed to move freely in and out of groups.

Our next goal then is to show that if a graph has an \epath\ $P$, then it has
one where \emph{few} edges are special. This is summarized in Lemma
\ref{lem:group} and Corollary \ref{cor:count}. Intuitively, the core idea is a
flipping argument: if the same group types appear too many times in a solution,
we can reverse a sub-path to obtain a solution with fewer groups.

\begin{lemma}\label{lem:group}

Let $G$ be an \egraph. Then, there exists an \epath\ $P$ of $G$ with the
following property: for any $i,j\in \{1,\ldots,k\}$, an edge of type $j$
appears directly after an edge of type $i$ at most once.  

\end{lemma}

\begin{proof}

Suppose that $P'$ is an \epath\ of $G$ in which there exist two edges of type
$i$, say $e_1^i,e_2^i$, and two edges of type $j$, say $e_1^j,e_2^j$ such that
$e_1^i$ is followed by $e_1^j$ in the path and $e_2^i$ is followed by $e_2^j$.
Without loss of generality, assume that $e_1^j$ appears before $e_2^i$ in the
path. We transform the path by reversing the order of all edges appearing
between $e_1^j$ and $e_2^i$ inclusive. In the new path $e_1^i$ is followed by
$e_2^i$, which is allowed, since they share a common endpoint ($u_i$).
Similarly, $e_1^j$ is followed by $e_2^j$.

Observe that the new path has strictly fewer groups. Therefore, repeating this
process at most a linear (in $|E|$) number of times we obtain an \epath\ $P$
with the stated property.  \qed     

\end{proof}

\begin{corollary}\label{cor:count}

Let $G$ be an \egraph. Then, there exists an \epath\ $P$ of $G$ such that for
all $i\in\{1,\ldots,k\}$, $P$ contains at most $(k-1)$ groups of type $i$.
Therefore, $P$ contains at most $k^2$ groups in total, and for each
$i\in\{1,\ldots,k\}$ there exist at most $2k$ special edges of type $i$.

\end{corollary}

We have now proved that if a solution exists, it must have a certain nice form.
Let us make one more easy observation.

\begin{lemma}\label{lem:large_groups}

Let $G(V,E)$ be an \egraph. Then, there exists an \epath\ $P$ such that, for
all $i\in\{1,\ldots,k\}$ for which there exist at least $k$ edges of type $i$,
$P$ has a group of type $i$ with  size at least $2$.  

\end{lemma}

\begin{proof}

By Corollary \ref{cor:count} there are at most $k-1$ groups of type $i$, so by
pigeonhole principle, one must contain at least two edges. \qed

\end{proof}

Let us note that Lemma \ref{lem:group}, Corollary \ref{cor:count} and Lemma
\ref{lem:large_groups} still hold even if $G$ is a hypergraph. We will make use
of this in the next section.

We are now ready to state the main reduction rule and prove its correctness.

\begin{lemma}\label{lem:red_rule}

Let $G(V,E)$ be a graph, and $S=\{u_1,\ldots,u_k\}$ a vertex cover of $G$ of
size $k$. Suppose that there exists an edge $(u_i,w)$ satisfying the following:

\begin{enumerate}

\item $w\notin S$

\item There are at least $k+1$ edges of type $i$ in $G$

\item For all $u_j\in S$ such that $(u_j,w)\in E$ we have $|(N(u_i)\cap
N(u_j))\setminus S|>4k$

\end{enumerate}

Then $G(V,E)$ has an \epath\ if and only if $G'(V,E\setminus\{(u_i,w)\})$ does.

\end{lemma}

\begin{proof} 

For the easy direction, suppose that $G'$ has an \epath\ $P'$. There are at
least $k$ edges of type $i$ in $G'$, so by Lemma \ref{lem:large_groups} at
least one group of type $i$ contains two or more edges.  Then, $(u_i,w)$ can
simply be inserted between two edges of this group to obtain an \epath\ for
$G$.

For the converse direction, suppose that $G$ has an \epath\ $P$.  Let $e_1,e_2$
be the edges appearing immediately before and after $(u_i,w)$ in $P$. If
$e_1,e_2$ share an endpoint, we can delete $(u_i,w)$ from $P$ and obtain a
valid solution for $G'$. Therefore, suppose they do not, and since they both
share an endpoint with $(u_i,w)$ we assume without loss of generality that
$e_1$ is incident on $u_i$ and $e_2=(u_j,w)$. (Observe that here we have used
the fact that $G$ is a graph, so the rest of our argument do not generalize to
hypergraphs).

We know now by the last condition that $N(u_j)\cap N(u_i)$ contains at least
$4k+1$ vertices of $V\setminus S$. Observe that, by Corollary \ref{cor:count},
there are at most $2k$ special edges of type $i$ and $2k$ special edges of type
$j$. Thus, there is a vertex of $(N(u_i)\cap N(u_j))\setminus S$, call it $z$,
such that $(u_i,z)$ and $(u_j,z)$ are not special.

Because $(u_i,z)$ is not special, the two edges appearing immediately before
and after it are both incident on $u_i$. Therefore, deleting $(u_i,z)$ still
leaves us with a valid edge-path. Similar reasoning can be used for $(u_j,z)$.
We construct a path $P'$ as follows: delete $(u_i,w), (u_i,z)$ and $(u_j,z)$
from $P$ and then insert $(u_i,z),(u_j,z)$ between $e_1$ and $e_2$. This is a
valid solution for $G'$. \qed

\end{proof}

Lemma \ref{lem:red_rule} now leads to the following theorem.

\begin{theorem}\label{thm:vckernel_main} 

\ehp\ has a kernel with $O(k^3)$ edges, where $k$ is the size of the input
graph's vertex cover.  

\end{theorem}

\section{Hypergraphs} \label{sec:hyper}

In this section we present an FPT algorithm for \ehp\ on hypergraphs
parameterized by the size of a (supplied) hitting set. As an interesting
consequence, our algorithm also establishes fixed-parameter tractability for a
novel parameterization of \textsc{Hamiltonian Path}, namely when the parameter
is the chromatic number of the input graph's complement.

In this section,  $G(V,E)$ will be a hypergraph (that is, $E$ is a collection
of arbitrary subsets of $V$). We assume that the input also contains a hitting
set $S\subset V$ of size $k$, that is, a set of vertices that intersects all
hyperedges.  Unlike the previous section, this is not an inconsequential
assumption, since finding even an approximate hitting set is generally a hard
problem.

We will rely on the fact that much of the material of the previous section
carries through unchanged.  In particular, Definitions \ref{def:type},
\ref{def:group}, also apply to hypergraphs.  Then, Lemma \ref{lem:group},
Corollary \ref{cor:count}, and Lemma \ref{lem:large_groups} hold for the case
of hypergraphs as well.  Unfortunately, Lemma \ref{lem:red_rule} does not seem
to generalize naturally in this case.

Let us thus describe a different algorithm for this problem. As mentioned, one
way to proceed is to try to identify the special edges, which form the backbone
of a path. Once these have been found, the problem becomes much easier.  We
will use a color-coding scheme to assist us in selecting these special
hyperedges.  The high-level idea is the following: for every
$i\in\{1,\ldots,k\}$ such that there are at least $2k$ hyperedges of type $i$,
color these hyperedges with $2k$ colors uniformly at random.  Then,
\emph{merge} (that is, take the union) of all hyperedges of type $i$ that took
the same color to a single hyperedge.  This process results in a hypergraph
$G'$ with $O(k^2)$ hyperedges.  We want to show that if this hypergraph has an
\epath\ then $G$ does as well, while if $G$ has an \epath\ then $G'$ has one
with non-negligible probability. The ``good colorings'' that give us this
non-negligible probability are those that assign a different color to each
special edge.

We are now ready to state the main result of this section.

\begin{theorem} \label{thm:hyper}

Given a hypergraph $G(V,E)$ and a  hitting set $S=\{u_1,\ldots,u_k\}$ of $G$,
there is an FPT algorithm that decides if $G$ has an \ehp\ in time
$2^{O(k^2)}n^{O(1)}$.

\end{theorem}

An interesting consequence of Theorem \ref{thm:hyper} is that it implies
fixed-parameter tractability for a non-standard parameterization of
\textsc{Hamiltonian Path}. The parameterization we are considering is by the
\emph{complement chromatic number}, that is, the chromatic number of the input
graph's complement. We are naturally led to this observation, because the line
graph of a hypergraph with a hitting set of size $k$ has a vertex set that can
be partitioned into at most $k$ cliques.  To the best of our knowledge, this
parameterization of \textsc{Hamiltonian Path} has not been considered before.

\begin{corollary}

Given a graph $G(V,E)$ and a proper $k$-coloring of its complement graph, there
exists an FPT algorithm that decides if $G$ has a Hamiltonian Path in time
$2^{O(k^2)}n^{O(1)}$.

\end{corollary}

\begin{proof}

The vertex set of $G$ can be partitioned into $k$ cliques. We will build a
hypergraph $G'(V',E')$ such that $G$ is the line graph of $G'$. It follows that
$G$ has a Hamiltonian Path if and only if $G'$ has an \epath.

We set $V'=\{1,\ldots,k\} \cup E$. For each $v\in V$ we use $I(v)$ to denote
the set of edges incident on $v$ and $c(v)$ to denote the color that $v$ has in
the given coloring. We set $E'= \{ I(v)\cup\{c(v)\}\ |\ v\in V\}$, or in other
words, we create a hyperedge for each vertex and include into it its incident
edges and its color. It is not hard to see that $G'$ has a hitting set of size
$k$ and that $G$ is the line graph of $G'$. \qed

\end{proof}

\section{Treewidth and Clique-width}

In this section we consider the \ehc\ problem parameterized by treewidth or
clique-width. As is customary for these parameters, we will assume that a
decomposition of width $k$ (or a clique-width expression with $k$ labels) is
given to us with the input. 

Let us first consider treewidth. One obvious approach we could try to follow is
to use the fact that if $G$ has treewidth $k$ its \emph{line} graph has
clique-width $O(k)$ (\cite{GW07}). Since deciding \ehc\ on $G$ is equivalent to
deciding \textsc{Hamiltonian Cycle} on its line graph, this would give an XP
algorithm, using known results for the latter problem (this is similar to the
approach of \cite{DDH14}).  Unfortunately, since \textsc{Hamiltonian Cycle} is
W-hard for clique-width, this approach could not lead to an FPT algorithm for
\ehc\ on treewidth. We thus have to recast the problem.

We will rely on Theorem \ref{thm:hn}, which states that the existence of an
\ecycle\ is equivalent to the existence of a \dgraph.  Thus, we can view \ehc\
as a subgraph problem, rather than an ordering problem.  This formulation
allows us to express the problem in a variant of MSO logic, without reference
to orderings.  We can then invoke standard meta-theorems to obtain
fixed-parameter tractability for treewidth.

Let us sketch the basic idea. Recall that MSO$_2$ logic allows one to express
properties involving sets of vertices \emph{or} edges (see \cite{CE12}). \des\
is the problem of looking for a set of vertices $V'$ and a set of edges $E'$
such that: all edges of $E$ have an endpoint in $V'$; the graph $G'(V',E')$ is
connected; all vertices of $G'(V',E')$ have even degree. The first two
properties are well-known to be expressible in MSO logic. Interestingly, the
third property is expressible in Counting MSO$_2$ (CMSO$_2$) logic, an
extension of MSO$_2$ which is still FPT for treewidth \cite{HOS08,C90}. Thus,
\ehc\ is expressible in CMSO$_2$ and is therefore FPT for treewidth.

We can use standard techniques to obtain the following:

\begin{theorem} \label{thm:tw}

Given a graph $G$ and a tree decomposition of width $k$, there exists an
algorithm deciding if $G$ has an \ecycle\ in time $k^{O(k)} n^{O(1)}$.

\end{theorem}

Let us now move to the main result of this section, which is the tractability
of \ehc\ parameterized by clique-width. Our high-level strategy will be to
eliminate complete bipartite subgraphs from the input graph, without increasing
the graph's clique-width and without affecting the answer of the problem. If we
can repeat this process we will in the end have a graph with small clique-width
and no large complete bipartite subgraphs. By Theorem \ref{thm:gw} the graph
will have small treewidth and we can use Theorem \ref{thm:tw}.  

Our main tool will be a reduction lemma (Lemma \ref{lem:reduce}). Roughly
speaking, the lemma states that if we find a sufficiently large complete
bipartite graph in $G$ with bipartition $A,B$, we can reduce it as follows:
first we remove all its edges and then we add three new vertices which are
connected to all vertices of both $A$ and $B$.  This transformation should not
affect the answer.

To prove Lemma \ref{lem:reduce} it will be useful to first prove the following
statement. Roughly speaking, it says that if a graph contains a $K_{3,3}$ (or
larger) complete bipartite subgraph then any \des\ can be edited to produce a
solution using all its vertices.

\begin{lemma} \label{lem:contain}

Let $G(V,E)$ be a graph and $A,B\subseteq V$, with $A,B$ disjoint sets, $|A|,
|B|\ge 3$ and $A\times B\subseteq E$. If $G$ has a \dgraph\ then it also has a
\dgraph\ $G_0(V_0,E_0)$ such that $(A\cup B)\subseteq V_0$ and $E_0\cap
(A\times B)\neq \emptyset$. 

\end{lemma}

\begin{proof}

Suppose that $G$ has a \dgraph\ $G_0(V_0,E_0)$. We will edit this solution by
adding vertices and adding or removing edges until the stated properties are
achieved.  In the remainder, when we say that we \emph{flip} an edge $e$ we
mean that, if $e\in E_0$ then we remove it from $E_0$, otherwise we add it to
$E_0$ and add its endpoints to $V_0$.

Let us first establish that $|V_0\setminus (A\cup B)|\le 1$ as follows: if
$V_0$ does not fully contain one of the two sets $A,B$, it must fully contain
the other (because $V_0$ is a vertex cover). Suppose without loss of generality
that $B\subseteq V_0$. If there exist $v_1,v_2\in A\setminus V_0$ then pick two
vertices $u_1,u_2\in B$.  We can flip all the edges of
$\{u_1,u_2\}\times\{v_1,v_2\}$ and produce a valid solution with more vertices.

Now, if there is a single vertex $v_1\in A\setminus V_0$ then we have two
cases: if there exist  $u_1\in B,v_2\in A$ such that $(u_1,v_2)\notin E_0$, we
pick an arbitrary $u_2\in B$ and flip the edges $\{u_1,u_2\}\times
\{v_1,v_2\}$. This produces a valid \dgraph\ that contains $v_1$. In the final
case, all edges of $A\times B$ not incident on $v_1$ are used in $E_0$. Then,
picking two arbitrary $u_1,u_2\in B$ and a vertex $v_2\in A$ and flipping the
edges $\{u_1,u_2\}\times \{v_1,v_2\}$ produces a valid solution that includes
$v_1$.  We can conclude that $A\subseteq V_0$. 

For the second property, observe that if $E_0$ does not use any edges of
$A\times B$ then we can add an arbitrary cycle to $E_0$ using edges of $A\times
B$ producing a valid solution. \qed

\end{proof}

\begin{lemma} \label{lem:reduce}

Let $G(V,E)$ be a graph and $A,B\subseteq V$ with $A,B$ disjoint sets,
$|A|,|B|\ge 5$ and $A\times B \subseteq E$. Let $C$ be a set of three new
vertices. Consider the graph $G'(V',E')$ where $V'=V\cup C$ and $E'=(E\setminus
A\times B)\cup (A\times C) \cup (B\times C)$. Then $G'$ has an \ecycle\ if and
only if $G$ does.

\end{lemma}

\begin{proof}

Let us first give a high-level description of the argument, which will be
expressed in terms of the \des\ problem.  Informally, it will be easy to
transform a solution for $G$ to one for $G'$, by replacing all edges of
$A\times B$ used in a \dgraph\ by paths of length 2 through the vertices of
$C$.  The more interesting part is the converse direction. Here, we will first
select appropriate edges of $A\times B$ to give all vertices even degrees in
$G$. The problem will be to do this in a way that ensures connectivity. For
this we will be needing the fact that we have a sufficiently large complete
bipartite graph.  Let us now give the details.

First, suppose that $G$ has a \dgraph\ $G_0(V_0,E_0)$.  We will now describe a
\dgraph\ $G_0'(V_0',E_0')$ of $G'$. We set $V_0'=V_0\cup C$, which is clearly a
vertex cover of $G'$. To construct $E_0'$ we begin with the set of edges
$E_0\setminus (A\times B)$.  Then, for each $(u,v)\in E_0\cap (A\times B)$ we
add to $E_0'$ the three distinct paths of length 2 that go from $u$ to $v$
through $C$. Observe that this process ensures that in the end all vertices of
$A,B$ have degree with the same parity as in $G_0$ and all vertices of $C$ have
even degree. The graph constructed is connected, because by Lemma
\ref{lem:contain} at least one edge of $A\times B$ is included in $E_0$.

For the converse direction, suppose we have a \dgraph\ $G_0'(V_0',E_0')$ of
$G'$.  By Lemma \ref{lem:contain}, because $C,(A\cup B)$ form two parts of a
sufficiently large complete bipartite subgraph we can assume that $ (A\cup
B\cup C)\subseteq V_0'$.  

We build a \dgraph\ $G_0(V_0,E_0)$ of $G$ as follows. First, $V_0=V_0'\setminus
C$, which is a vertex cover of $G$.   Let $E_C$ be the set of edges of $E_0'$
incident on $C$.  It must be the case that $|E_C|$ is even, since all vertices
of $C$ have even degree in $G_0'$ and $C$ is an independent set. We start
building $E_0$ by including all the edges of $E_0'\setminus E_C$. We will now
go through two phases of ``fixing'' $E_0$ by adding to it edges of $A\times B$.

Initially, we concentrate on making all degree parities even. We will say that
we \emph{flip} an edge $e$ to mean that, if $e\in E_0'$ then we remove it from
$E_0'$, otherwise we add it to $E_0'$.  Observe that, for our current selection
of $E_0'$, the number of vertices of $A\cup B$ with odd degree is even. This is
a consequence of the fact that $|E_C|$ is even and that all vertices have even
degrees in $G_0'$. As long as there exist two vertices of $A\cup B$ with odd
degree, select a shortest path connecting them and flip its edges. Repeating
this will eventually produce a set $E_0$ that makes the degree of all vertices
even.

We now need to augment $E_0$ to make sure that $G_0$ is also connected. It is
not hard to see that if $G_0$ is not connected there must be two vertices of
$A\cup B$ in different components (otherwise, we could find a disconnected
component in $G_0'$). 

Suppose that for one of the sets, say $A$, there exist two vertices $v_1,v_2\in
A$ such that $v_1,v_2$ are in different components. Clearly, their
neighborhoods $N(v_1),N(v_2)$ must be disjoint. At the same time, if $v_1,v_2$
have two common non-neighbors $u_1,u_2\in B$, we can add the edges of
$\{u_1,u_2\}\times\{v_1,v_2\}$ to $E_0$ and obtain a valid solution with fewer
components. Thus, it must be the case that $N(v_1),N(v_2)$ cover all of $B$,
except for at most one vertex. Because of the size of $B$ this means that for
one of them, say $v_1$, we have $|N(v_1)\cap B|\ge 2$.

Consider now an arbitrary $v_3\in A$. Clearly, it either has no neighbors in
$N(v_1)$ or its has no neighbors in $N(v_2)$ (otherwise $v_1,v_2$ would be in
the same component). If it has no neighbors in $N(v_1)$ then we add all edges
between $\{v_2,v_3\}$ and two arbitrary vertices of $N(v_1)$ to improve the
solution. Therefore, every vertex of $A$ except $v_2$ has some neighbor in
$N(v_1)$. Thus, by the above steps we have made sure that, if $A$ is not
contained in a single component, then there exists a component that contains
all but one of the vertices of $A$.

Let $S$ be the component that contains almost all the vertices of $A$. If there
are two vertices $u_1,u_2\in B\setminus S$ then $u_1,u_2$ have two common
non-neighbors in $S$ and we can again augment the solution. Thus, $S$ also
contains $B$, except for at most one vertex.

We are now almost done. If there exists $v_2\in A\setminus S$ we can handle it
as follows. If there are $v_1\in A\cap S$, $u_1\in B\cap S$ such that
$(v_1,u_1)\notin E_0$ then select an arbitrary vertex $u_2\in B$ and flip the
edges of $\{u_1,u_2\}\times\{v_1,v_2\}$. This improves the solution by
including $v_2$ in $S$. If on the other hand all edges of $(A\cap S)\times
(B\cap S)$ are in $E_0$ we can select arbitrary $v_2\in A\cap S$ and
$u_1,u_2\in B$ and flip the edges of $\{u_1,u_2\}\times\{v_1,v_2\}$. Because
$|B\cap S|\ge 3$ and $|A\cap S|\ge 2$ this will strictly increase the component
$S$.  A symmetric argument can handle the possible remaining vertex of $B$.
\qed

\end{proof}

We are now almost ready to proceed with our algorithm. To simplify
presentation, we will only apply Lemma \ref{lem:reduce} to subgraphs which are
at least as large as $K_{7,7}$. Observe that in such a case, $G'$ has strictly
fewer edges than $G$. It is then clear that the reduction is making progress
and after a bounded number of applications  we get a graph with no large
complete bipartite subgraphs.

There is, however, one problem that remains. We must also show that we can
apply Lemma \ref{lem:reduce} repeatedly without increasing the graph's
clique-width. If we cannot guarantee this, then, even though we will have
eliminated large $K_{t,t}$ subgraphs, we will not be able to invoke Theorem
\ref{thm:gw} in the end. We therefore have to take care to only apply the
reduction rule in some specific situations. For this, we will have to work with
the given clique-width expression of $G$.

Our first step is to handle an obvious part of the given clique-width
expression where large bipartite subgraphs are constructed, namely, the join
operation.

\begin{lemma} \label{lem:bigjoin}

Given a graph $G$ and a clique-width expression with $k$ labels it is possible
to produce in polynomial time a graph $G'$ and a clique-width expression with
$k+2$ labels such that:

\begin{enumerate}

\item $G$ has an \ecycle\ if and only if $G'$ does

\item For every join operation in the expression of $G'$, one of the two
involved sets of vertices contains at most $6$ vertices.

\end{enumerate}

\end{lemma}

Unfortunately, Lemma \ref{lem:bigjoin} is not enough to guarantee the
elimination of large complete bipartite subgraphs, since these may also be
constructed gradually. However, eliminating big joins gives our clique-width
expression a certain structure which we can leverage to deal with the remaining
bi-cliques efficiently.

\begin{lemma} \label{lem:bicliques}

Given a graph $G(V,E)$ and a clique-width expression with $k$ labels and the
property that for all join operations one involved set has size at most $6$, we
can in polynomial time produce a graph $G'$ with clique-width $k+2$ such that
$G'$ does not contain $K_{21k,21k}$ as a subgraph.

\end{lemma}

We can now describe our algorithm. Given a graph $G$ and a clique-width
expression with $k$ labels, we first invoke the algorithms of Lemmata
\ref{lem:bigjoin},\ref{lem:bicliques}. We are thus left with a graph with
clique-width at most $k+4$ and no complete bipartite subgraph larger than
$K_{t,t}$ for $t=O(k)$. By Theorem \ref{thm:gw}, this graph has treewidth
$O(k^2)$. We can now apply an FPT algorithm to obtain a reasonable tree
decomposition (see e.g.~\cite{BDD13}) and then invoke Theorem \ref{thm:tw}.

\begin{theorem}

Given a graph $G$ and a clique-width expression with $k$ labels, there exists
an algorithm that decides if $G$ has an \ecycle\ in time $k^{O(k^2)} n^{O(1)}$.

\end{theorem}

\bibliographystyle{abbrv}
\bibliography{ehp}
\newpage
\appendix

\section{Omitted Proofs}

\subsection{Proof of Lemma \ref{lem:equiv}} 

\begin{proof}

Suppose that \ehc\ is FPT for one of these parameters. We are given a graph
$G(V,E)$ where we want to decide \ehp. Let $u,v\in V$ and let $G'$ be obtained
from $G$ by adding a path of length 3 (through new vertices) from $u$ to $v$.
It is not hard to see that $G'$ has an \ehc\ if and only if $G$ has an \ehp\
where $u$ appears in the first edge and $v$ in the last. Repeating this process
for all pairs of vertices $u,v\in V$ allows us to decide \ehp\ on $G$. Observe
that for all considered parameters their values are only changed by an additive
constant.

For the converse direction, suppose that we have an FPT algorithm for \ehp\ and
we want to decide \ehc\ on $G(V,E)$. Select a vertex $u\in V$ and attach to it
two distinct paths of length two (through new vertices). The new graph has an
\ehp\ (which must start and end at the endpoints of the new paths) if and only
if $G$ has an \ehc\ where the first and last edges both include $u$. Trying all
possibilities for $u$ lets us decide \ehc\ on $G$. Again the parameters are not
affected by more than an additive constant. \qed

\end{proof}

\subsection{Proof of Theorem \ref{thm:vckernel_main}} 

\begin{proof}

The algorithm is simple: as long as there exists an edge $(u_i,w)$ for which
the conditions of Lemma \ref{lem:red_rule} apply, delete this edge. This can be
done in polynomial time. We will show that, once we can no longer apply this
reduction, the graph has the promised size.  To prove this, for each vertex
$u_i\in S$, we will we show that the number of edges of type $i$ incident on
$V\setminus S$ is at most $4k^2$.  Observe that the theorem then immediately
follows.

Suppose that, for some $i\in\{1,\ldots,k\}$, there exist $4k^2$ edges of type
$i$ incident on $V\setminus S$.  As a first step, note that if $N(u_i)\setminus
S$ contains any vertices of degree 1,  we can apply Lemma \ref{lem:red_rule},
because for such vertices the last condition is vacuously true. Suppose then
that all vertices of $N(u_i)\setminus S$ have another neighbor in $S$.

We say that $u_j$ has a \emph{small} overlap with $u_i$ if $|(N(u_i)\cap
N(u_j))\setminus S|\le 4k$. All vertices of $N(u_i)\setminus S$ satisfy the
first condition of Lemma \ref{lem:red_rule}, while $i$ satisfies the second
one. Thus, to prove that we can still apply the rule we only need to find a
vertex of $N(u_i)\setminus S$ such that all its neighbors in $S$ have large
overlap with $u_i$.

There are at most $k-1$ vertices in $S$ that have small overlap with $u_i$.
These have at most $4k(k-1)$ neighbors in $N(u_i)\setminus S$. Thus, if this
set has size $4k^2>4k(k-1)$, there must exist an edge to which we can apply the
reduction rule, because its endpoint in $V\setminus S$ only has neighbors with
a large overlap with $u_i$.  \qed

\end{proof}

\subsection{Proof of Theorem \ref{thm:hyper}} 

\begin{proof}

We will first describe a randomized color-coding algorithm that achieves the
promised result. In the end, we also explain how this algorithm can be
derandomized with standard techniques.

Let us describe the algorithm more formally. For each $i\in\{1,\ldots,k\}$ let
$E_i$ be the set of hyperedges of type $i$. If $|E_i|>2k$ then do the
following: Randomly select $2k$ hyperedges of type $i$ and color each with a
distinct color from $\{1,\ldots,2k\}$. Then color all remaining hyperedges of
type $i$ uniformly at random with a color from $\{1,\ldots,2k\}$. Note that
this process ensures that all colors are used at least once, which will
simplify some arguments.

Let $E_i^c$ be the set of hyperedges of type $i$ that received color $c$.  Now
for each colored set $E_i^c$ construct a new hyperedge $e_{i,c} = \cup_{e\in
E_i^c} e$.  Remove all hyperedges of $E_i^c$ from $G$ and replace them with the
new hyperedge $e_{i,c}$. After performing this process for all
$i\in\{1,\ldots,k\}$ the hypergraph has at most $2k^2$ hyperedges. We then use
an exponential-time algorithm to solve \ehp\ on this new graph in time
$2^{O(k^2)}$. If the new graph $G'$ has an \ehp\ we decide that $G$ also does,
otherwise we reply that it does not.

To show that the problem can be solved with the above procedure we will
establish two properties:

\begin{itemize} 

\item if $G$ has an \epath\ $P$, then there exists an \epath\ $P'$ for the new
hypergraph $G'$ with probability at least $e^{-2k^2}$; 

\item if the new hypergraph $G'$ has an \epath\ $P'$, then there exists an
\epath\ $P$ in the original graph.  

\end{itemize}

Observe that if we achieve the above, a randomized FPT algorithm which
correctly decides the problem follows by simply repeating this process a
sufficiently large number of times. Let us therefore establish these
properties.

For the first direction, assume that $G$ has an \epath\ $P$ and (by Corollary
\ref{cor:count}) there are at most $2k$ special hyperedges of each type. We say
that the coloring of the edges of $E_i$ is \emph{good} if all the special
hyperedges of type $i$ received different colors. The probability that the
coloring of $E_i$ is good is at least $\frac{(2k)!}{(2k)^{2k}} > e^{-2k}$. The
probability that all edge types which were randomly colored received a good
coloring is therefore at least $e^{-2k^2}$. We can now show that if the
coloring is good for all edge types then $G'$ has an \epath. Start with $P$.
If $P$ contains two hyperedges of the same type and color, one of them is not
special (because the coloring is good). Delete the non-special hyperedge from
the path.  The path is still valid, since the two neighbors of the deleted
hyperedge share a common endpoint.  Repeat this until in the end we are left
with a single hyperedge of each color.  Replace the remaining hyperedge of type
$i$ that received color $c$  with the hyperedge $e_{i,c}$ of $G'$.  Doing this
for each type $i$ that was randomly colored produces a valid \epath\ of $G'$.

For the converse direction, suppose we have an \epath\ $P'$ of $G'$. We will
first build from this a valid edge-path of $G$, and then insert into it the
remaining hyperedges to obtain an \epath. For the first step, as long as $P'$
contains one of the new hyperedges $e_{i,c}$ do the following: find a vertex
$v_1$ that is common between $e_{i,c}$ and the hyperedge that precedes it and a
vertex $v_2$ that is common with the hyperedge that follows. It must be the
case that some hyperedge of type $i$ and color $c$ contains $v_1$, call it
$e_1$. Similarly, some hyperedge (not necessarily distinct from $e_1$) contains
$v_2$, call it $e_2$. Replace the hyperedge $e_{i,c}$ with $e_1,e_2$ (or just
$e_1$ if they are the same hyperedge). This is still a valid edge-path, so
repeating this process gives a valid edge-path made up of original hyperedges
of $G$. Let $E_s$ be the set of hyperedges of this path.

By definition, the graph $G''(V,E_s)$ contains an \epath.  Recall now that for
all $i$ that were randomly colored and all colors $c$, $G'$ contained a
hyperedge $e_{i,c}$, which has now been replaced by one or two hyperedges of
type $i$ in $E_s$. This means that $E_s$ contains at least $2k$ hyperedges of
type $i$.  By Lemma \ref{lem:large_groups}, $G''(V,E_s)$ has an \epath\
containing a group of type $i$ with at least two hyperedges. Take all
hyperedges of $E_i\setminus E_s$ and insert them between two hyperedges of that
group.  Repeating this process produces an \epath\ of $G$. 

Finally, let us sketch how the above algorithm can be derandomized. The
important point of this analysis is that there exist at most $2k^2$ special
edges for which we hope to use distinct colors. Rather than coloring each type
independently then, we could color all affected hyperedges with colors from
$\{1,\ldots,2k^2\}$.  It is now sufficient to try a set of colorings such that
any set of $2k^2$ hyperedges becomes colorful for some coloring. As is standard
in these situations, we can use a perfect hash function family from
$\{1,\ldots,|E|\}$ to $\{1,\ldots,2k^2\}$. There exist such families with size
$2^{O(k^2)}\log |E|$ (\cite{AYZ95,SS90}).  \qed

\end{proof}

\subsection{Proof of Theorem \ref{thm:tw}} 

\begin{proof}

We only sketch the algorithm, since it follows the usual treewidth dynamic
programming pattern.  We follow the conventions of \cite{BK08}. For each node
$i$ of a nice tree decomposition let $G_i$ be the subgraph of $G$ induced by
vertices appearing in the bags of the sub-tree rooted at $i$. We define a
dynamic programming table $C_i$ that characterizes a partial solution when
restricted to the graph $G_i$.  If $X_i$ is the set of vertices contained in
the bag $i$, then $C_i$ is a set of triples $(S,R,P)$ where $P\subseteq
S\subseteq X_i$, and $R$ is an equivalence relation on $S$ (i.e.~a partition of
$S$). The intuitive meaning is the following: $S$ contains the vertices of the
bag which have been selected as part of the \des\ (and therefore must form part
of a vertex cover of the graph).  We use $P$ to remember which vertices of $S$
have an odd number of edges incident on them selected. In addition, we use $R$
to remember which vertices of $S$ are in the same connected component, in the
graph constructed using already selected edges. 

More formally, we want to make sure that a triple $(S,R,P)$ belongs in $C_i$ if
and only if there exists a subgraph $G_i'(V_i,E_i)$ of $G_i$ such that:

\begin{enumerate}

\item $V_i$ is a vertex cover of $G_i$ and $V_i\cap X_i = S$

\item For all $u,v\in S$ we have $uRv$ if and only if $u$ is reachable from $v$
in $G_i'$.  Furthermore, all vertices of $V_i\setminus X_i$ are reachable from
some vertex of $S$ in $G_i'$

\item For all $u\in S$ we have $u\in P$ if and only if $u$ has odd degree in
$G_i'$. Furthermore, all $u\in V_i\setminus X_i$ have even degree in $G_i'$

\end{enumerate}

Given the above description, the dynamic programming table for each node can be
computed using standard techniques: we just need to make sure that $S$ is a
vertex cover, and that we never ``forget'' a vertex with odd degree or the last
vertex of a component.  In the end, we check if the root contains an entry
$(S,\{S\},\emptyset)$ for some set $S$. Notice that the running time is
dominated by the size of the dynamic programming tables, which are in turn
dominated by the number of partitions of $S$. This is upper-bounded by the
$k$-th Bell number, which is asymptotically less than $k^k$. \qed

\end{proof}

\subsection{Proof of Lemma \ref{lem:bigjoin}} 

\begin{proof}

We will use two new labels $k+1,k+2$. Informally, the first is a ``work'' label
and the second a ``garbage'' label. Given the clique-width expression of $G$ we
can identify in polynomial time a large join operation.  Suppose that there is
an operation joining labels $i,j$ and the sets $V_i,V_j$ of vertices with the
corresponding labels have size at least $7$.

Remove the offending join operation and replace it with the following
operations: introduce $3$ new vertices with label $k+1$, join labels $i$ and
$k+1$, join labels $j$ and $k+1$, rename label $k+1$ to $k+2$.  

The fact that the anwer to the \ehc\ problem does not change follows directly
from Lemma \ref{lem:reduce}. Repeating the above process eliminates all large
joins in polynomial time. \qed

\end{proof}

\subsection{Proof of Lemma \ref{lem:bicliques}} 

\begin{proof}

As mentioned, we view the given clique-width expression as a rooted binary
tree. Given a node $x$ of that tree, $G_x(V_x,E_x)$ is the graph corresponding
to the clique-width sub-expression rooted at $x$. 

Consider a graph $G_x$ and the set of vertices with label $i$ in $G_x$, call it
$V_i^x$. If there also exists a set $B\subseteq V\setminus V_x$ such that
$|B|,|V_i^x|\ge 7$ and $B\times V_i^x \subseteq E$ we apply a simplifying
transformation. Specifically, immediately after the construction of $G_x$ we
insert the following operations: introduce $3$ vertices with label $k+1$, join
$i$ to $k+1$, rename $i$ to $k+2$, rename $k+1$ to $i$.

Correctness of the above transformation again follows from Lemma
\ref{lem:reduce}. The important point here is that we can set all the vertices
of $V_i^x$ to the ``garbage'' label $k+2$ and allow them to be ``represented''
by the $3$ new vertices. Any vertex of $V\setminus V_x$ that had an edge to a
vertex of $V_i^x$ had an edge to all of them in $G$. Such vertices can
therefore be assumed to belong to $B$. These vertices will be connected to the
three newly introduced vertices. Observe also that this procedure can be
carried out in polynomial time, since if we fix one side of a complete
bipartite subgraph (in this case $V_i^x$) it is easy to find the maximum $B$ in
$G$.

What remains to argue is that repeated applications of the above procedure will
necessarily remove \emph{all} large bi-cliques. Equivalently, we need to prove
that if $G$ has a large bi-clique then there exists a $V_i^x$ to which the
above reduction rule applies. We will also use the fact that no large joins
exist in the clique-width expression.

Suppose that a graph $G_x$ corresponding to some sub-expression contains
$K_{t,t}$, $t\ge 14k$ as a subgraph on the sets of vertices $A_x,B_x$. We claim
that there exists a descendant $y$ of $x$ such that $G_y$ contains $K_{t',t'}$
as a subgraph on the sets $A_y\subseteq A_x, B_y\subseteq B_x$, with $t-7k\le
t' < t$ and $t'$ maximal. To see this, observe that if the claim were not true,
the two closest disjoint descendants of $G_x$, call them $G_y,G_z$, that
contain vertices of $A_x,B_x$ would both contain at least $7k$ vertices. This
would mean (without loss of generality) that $G_y$ would contain $7$ vertices
of $A_x$ having the same label and $G_z$ would contain $7$ vertices of $B_x$
having the same label.  But, since we disallow large joins it would be
impossible to construct the edges joining these vertices in $G_x$.

Suppose now that $G$ contains a $K_{21k,21k}$ on the sets $A,B$. By repeated
application of the above claim there is a subgraph $G_x$ containing a $K_{t,t}$
on sets $A_x\subseteq A$, $B_x\subseteq B$, where $7k\le t \le 14k$ and $t$ is
maximal.  Consider the larger of the two sets $A_x,B_x$, say $A_x$. It must
contain $7$ vertices with the same label, call this set $V_i^x$. On the other
hand there are at least $7k\ge 7$ vertices in $B\setminus B_x$, which will
eventually all be joined to $V_i^x$. We have thus found a set to which we could
apply our reduction rule.  \qed 

\end{proof}

\end{document}